\documentclass[aps,prx,showpacs,twocolumn,amsmath,amssymb,nofootinbib]{revtex4-1}
\usepackage{graphicx}
\usepackage{dcolumn}
\usepackage{bm}
\usepackage{amsmath}
\usepackage{amsfonts}
\usepackage{amssymb}
\usepackage{multirow}

\usepackage{amsthm}
\usepackage{pinlabel}
\usepackage{natbib}

\usepackage{epstopdf}
\usepackage[abs]{overpic}
\usepackage[dvipsnames]{xcolor}

\usepackage{tikz}
\usetikzlibrary{arrows.meta}
\usetikzlibrary{decorations.pathmorphing}
\usetikzlibrary{decorations.markings}

\tikzset{
	particle/.style={thick,draw=blue, postaction={decorate},
		decoration={markings,mark=at position .5 with {\arrow[blue]{triangle 45}}}},
	gluon/.style={decorate, draw=black,
		decoration={coil,aspect=0}}
}

\usepackage{placeins}
\usepackage{cancel}

\usepackage[dvipsnames]{xcolor}

\usepackage{braket}

\definecolor{specialgray}{HTML}{505050}
\definecolor{col10K}{HTML}{FFA000}
\definecolor{col300K}{HTML}{924FA4}
\definecolor{colMu}{HTML}{5278BD}
\definecolor{colMuI}{HTML}{924FA4}

\definecolor{specialgray}{HTML}{505050}
\definecolor{col10K}{HTML}{FFA000}
\definecolor{col300K}{HTML}{924FA4}
\definecolor{colMu}{HTML}{5278BD}
\definecolor{colMuI}{HTML}{924FA4}
\definecolor{newred}{HTML}{D53E4F}
\definecolor{newblue}{HTML}{5278BD}
\definecolor{newcyan}{HTML}{1EA0A0}
\definecolor{newgreen}{HTML}{5CB14E}
\definecolor{newpurple}{HTML}{924FA4}
\definecolor{newyellow}{HTML}{D1C72E}
\definecolor{neworange}{HTML}{D6923C}

\begin{document}

\title{
 Phonon-mode specific contributions to room-temperature superconductivity in atomic hydrogen at high pressures}
 \author{Ashok K. Verma}\email{hpps@barc.gov.in}
\author{P. Modak}
\affiliation{High Pressure and Synchrotron Radiation Physics Division, Bhabha Atomic Research Centre, Mumbai 400085, India}
\author{Fabian Schrodi}\email{fabian.schrodi@physics.uu.se}
\author{Alex Aperis}\email{alex.aperis@physics.uu.se}
\author{Peter M. Oppeneer}
\affiliation{Department of Physics and Astronomy, Uppsala University, P.\ O.\ Box 516, SE-75120 Uppsala, Sweden}

\vskip 0.4cm
\date{\today}

\begin{abstract}
	\noindent 
	We investigate the role of specific phonon mode symmetries for the room temperature superconductivity in atomic hydrogen under large pressure. Using anisotropic Migdal-Eliashberg theory with \textit{ab initio} input from density functional theory, we show that the $E_u$ phonon modes are the dominant driving force for obtaining such high critical temperatures. When going from 400 to 600 GPa, we find an increased transition temperature, however, the total electron-phonon coupling strength is counterintuitively reduced. Our analysis reveals that this is due to an enhanced contribution to the coupling strength by the $E_u$ phonon mode.
\end{abstract}

\maketitle

\section{Introduction}\label{scIntroduction}

Reaching superconductivity at room temperature has {been the} focus of intense research activities in the last few years (see \cite{Pickard2020,Flores2020} for recent surveys). Especially promising results have been {achieved for} superhydrides,
such as H$_3$S, with a transition temperature of 203 K at a pressure of 155 GPa  \cite{Drozdov2015},
LaH$_{10}$ with a $T_c$ around 250 K  at a pressure of 170 GPa or higher \cite{Liu2017,Drozdov2019,Somayazulu2019}, and YH$_{6}$ with {$T_c\simeq220\,\mathrm{K}$} at 166 - 237 GPa \cite{Troyan2019,Kong2019}. 
Very recent studies report room-temperature superconductivity (287 K) in a carbonaceous sulfur hydride at 267 GPa \cite{Snider2020}, and possibly even {a} higher {critical temperature} in a La superhydride mixed with ammonia borane \cite{Grockowiak2020}.
A unifying aspect of these recently discovered high-temperature superconductors is the prevalent conventional electron-phonon mechanism that is responsible for the record high critical temperatures \cite{Flores2020}.

The quest for room-temperature superconductivity in hydrides goes back to a proposal by Ashcroft \cite{Ashcroft1968}, stating that dense metallic atomic hydrogen could exhibit superconductivity at a very high critical temperature. The existence of {a} metallic {phase of} atomic hydrogen was first conceived by Wigner and Huntington in 1935 \cite{Wigner1935}. Since these seeding works, massive efforts have been devoted to {experimentally confirm such predictions at high pressures (see \cite{Mao1994,McMahon2012RMP}), eventually aiming for the final demonstration of high temperature superconductivity in this material. However, the formation of metallic atomic hydrogen at high pressure has been difficult to establish in diamond-anvil pressure cells. So far, some evidence for a metallic phase has been presented at various pressures, from 250 GPa to 495 GPa \cite{Hemley1989,Goncharov2001,Eremets2016,Dias2017,Ji2019,Loubeyre2020}, but the findings of these works are not yet unambiguously accepted by the entire scientific community.

To understand better the formation of superconductivity in hydrides at high transition temperatures, theory can provide valuable insight. It is widely accepted that the conventional electron-phonon mechanism is at play, being enhanced by the small ionic mass of hydrogen, the large electron-ion Coulomb interaction, and relatively weak electron-electron interaction. Although the appearance of superconductivity has not yet been reported, first-principles crystal structure investigations have determined that atomic hydrogen will adopt the $I4_1/amd$ structure for a large pressure interval of $500 - 1000$ GPa \cite{Pickard2007,McMahon2011,Degtyarenko2016}. Advanced quantum Monte-Carlo calculations estimated the transition pressure of 374 GPa for the transition from the molecular phase to the atomic $I4_1/amd$ phase \cite{Azadi2014}. Superconductivity in the latter phase has been investigated using first-principles electronic structure calculations of the electron and phonon bands and their coupling, using the semi-empirical McMillan and Allen-Dynes equation \cite{McMahon2011,McMahon2012err,Yan2011} or by solving the isotropic Eliashberg equations \cite{Durajski2014,Borinaga2016}.  The obtained transition temperatures $T_c$ are around room temperature for a Coulomb pseudopotential value $\mu^{\star} =0.10$.

In this work we present a phonon-mode resolved analysis of metallic hydrogen in the superconducting state at pressures of $400$ and $600\,\mathrm{GPa}$, where the $I4_1/amd$ phase is prevalent. Our calculations are carried out with the Uppsala Superconductivity (\textsc{uppsc}) code \cite{UppSC,Aperis2015,Schrodi2018,Schrodi2019,Schrodi2020_2,Schrodi2020_3,Schrodi2020_4}. Specifically, we solve here the anisotropic Migdal-Eliashberg equations using first-principles electron energies, phonon frequencies, and electron-phonon couplings as input. The total electron-phonon coupling constant $\lambda\simeq2.32$ at $400\,\mathrm{GPa}$ contains dominant contributions from the $B_{1g}$ phonon mode, while the $A_{2u}$ mode has the smallest impact. The remaining $E_u$ and $E_g$ modes both contribute with comparable and substantial magnitude to $\lambda$. We find $T_c$ approximately as room temperature for a reasonable range of Coulomb pseudopotential values $\mu^{\star}$, which is consistent with previous investigations \cite{McMahon2012err,Durajski2014,Borinaga2016}. Selectively investigating each of the phonon modes reveals that the $E_u$ mode contributes most to the $T_c$, despite having a subdominant role concerning the electron-phonon coupling strength. We provide a further proof of this observation by increasing the pressure to $600\,\mathrm{GPa}$, where the critical temperature slightly increases, despite a reduction in electron-phonon coupling strength $\lambda\simeq2.09$. 
In accordance with  the just described picture, our mode-resolved Eliashberg calculations reveal that this stems from an enhanced contribution from the $E_u$ mode.


\section{Methodology}\label{scTheory}

\subsection{First-principles calculations} 
We {perform} first-principles calculations within the density functional theory (DFT) framework  using the Quantum Espresso package \cite{Giannozzi2009}. We {adopt} the $I4_1/amd$ crystal structure of atomic hydrogen that was predicted to be the stable structure over a large pressure range of 400 to 1000 GPa \cite{McMahon2011}.  The exchange-correlation energy functional {is} treated within the generalized gradient corrected scheme of Perdew-Burke-Ernzerhof \cite{Perdew1996}. The interactions between valence electrons and core {are} treated within the projector-augmented-wave (PAW) approach and {the} plane wave basis set {is} constructed using an energy cut-off of 80 Ry. The Brillouin zone (BZ) integrations {are} carried out using a uniform dense $24\times24\times24$ Monkhorst-Pack {$\mathbf{k}$}-point grid. The phonon dispersions and electron-phonon couplings {are} calculated on a dense $12\times12\times12$ {$\mathbf{q}$}-point grid using density functional perturbation theory (DFTP).  All free parameters of the crystal lattice {are} optimized at 400 and 600 GPa. Since anharmonicity shows a minor effect on the critical temperature \cite{Borinaga2016}, these effects are not considered in the present calculations.

\subsection{Eliashberg theory calculations}

From {\em ab initio} calculations we obtain branch $\nu$ and wave vector $\mathbf{q}$ dependent phonon frequencies $\omega_{\mathbf{q},\nu}$, as well as electron-phonon coupling constants $\lambda_{\mathbf{q},\nu}$ and quasiparticle lifetimes $\gamma_{\mathbf{q},\nu}$. By defining bosonic Matsubara frequencies $q_l=2\pi Tl$, $l\in\mathbb{Z}$, at temperature $T$ we obtain the dynamic electron-phonon couplings via
\begin{align}
\lambda_{\mathbf{q},l} = \sum_{\nu} \lambda_{\mathbf{q},\nu} \frac{\omega_{\mathbf{q},\nu}^2}{\omega_{\mathbf{q},\nu}^2 + q_l^2} ~. \label{lambdaql}
\end{align}
In the above we use the notation $g(\mathbf{q},iq_l)=g_{\mathbf{q},l}$ for any function $g$ for the sake of brevity. The couplings calculated from Eq.\,(\ref{lambdaql}) serve as input for the self-consistent anisotropic Eliashberg equations
\begin{align}
Z_{\mathbf{k},m} &= 1 + \frac{\pi T}{\omega_m} \sum_{\mathbf{k}',m'} \frac{\delta(\xi_{\mathbf{k}'})}{N_0} \lambda_{\mathbf{k}-\mathbf{k}',m-m'} \frac{\omega_{m'}}{\sqrt{\omega_{m'}^2 + \Delta_{\mathbf{k}',m'}^2}} ~, \label{z} \\
\Delta_{\mathbf{k},m} &= \frac{\pi T}{Z_{\mathbf{k},m}} \sum_{\mathbf{k}',m'} \frac{\delta(\xi_{\mathbf{k}'})}{N_0} \left[ \lambda_{\mathbf{k}-\mathbf{k}',m-m'} - \mu^{\star}(\omega_c)\right] \nonumber\\
& ~~~~~~~~~~ \times  \frac{\Delta_{\mathbf{k}',m'}}{\sqrt{\omega_{m'}^2 + \Delta_{\mathbf{k}',m'}^2}} , \label{delta}
\end{align}
describing the electron mass renormalization $Z_{\mathbf{k},m}$ and superconducting gap function $\Delta_{\mathbf{k},m}$ \cite{Aperis2015}. Again we write $f(\mathbf{k},i\omega_m)=f_{\mathbf{k},m}$, now with fermion Matsubara frequencies $\omega_m=\pi T(2m+1)$, $m\in\mathbb{Z}$. We use $\mu^{\star}$ as Anderson-Morel Coulomb pseudopotential, which enters Eq.\,(\ref{delta}) with a Matsubara frequency cutoff $\omega_c$. The critical temperature $T_c$ is defined as the smallest $T$ at which the self-consistent solution to Eqs.\,(\ref{z}-\ref{delta}) yields a vanishing superconducting gap.

The electron density of states $N_0$ at the Fermi level is calculated via the adaptive smearing method, namely
\begin{align}
N_0 = \sum_{\mathbf{k},n} \frac{1}{\sqrt{2\pi}} \frac{1}{W_{\mathbf{k},n}} \exp\Big( -\frac{\xi_{\mathbf{k},n}^2}{2W_{\mathbf{k},n}^2} \Big) ~, \label{N0}
\end{align}
where the broadening tensor is defined as
\begin{align}
W_{\mathbf{k},n} = a\cdot\Delta k\cdot  \Big|  \frac{\partial \xi_{\mathbf{k},n}}{\partial \mathbf{k}} \Big|  ~, \label{wsmear}
\end{align}
in combination with the Methfessel-Paxton scheme\,\cite{Methfessel1989}. In Eq.\,(\ref{wsmear}), $\Delta k$ is the momentum resolution and $a$ can be chosen $\mathcal{O}(1)$\,\cite{Yates2007}. Furthermore, $\xi_{\mathbf{k},n}$ is the electron dispersion as computed from DFT, with $\mathbf{k}$ a Brillouin zone (BZ) momentum and $n$ a band index. We consider here only electronic states at the Fermi level, hence our calculations are carried out for the two partially occupied energy bands (shown further below).

We obtain a more simplified estimate of $T_c$ by employing the semi-empirical McMillan equation\,\cite{McMillan1968}, including a modification due to Allen and Dynes \cite{Allen1975},
\begin{align}
T_c = \frac{\omega_{\mathrm{\log}}}{1.2} \exp\Big( \frac{-1.04(1+\lambda)}{\lambda (1 - 0.62\mu^{\star}) - \mu^{\star}}\Big) ~.
\end{align}
Here $\lambda$ is the total electron-phonon coupling constant,
\begin{align}
\lambda &=\sum_{\mathbf{q},\nu}\lambda_{\mathbf{q},\nu} \label{lambda_sum} \\
& = 2\int_0^{\infty} \frac{\alpha^2F(\omega)}{\omega} \mathrm{d}\omega \label{lambda_int} ,
\end{align} 
and $\alpha^2 F(\omega)$ is  the real-frequency $\omega$ dependent Eliashberg function, given as
\begin{align}
\alpha^2F(\omega) = \frac{1}{2\pi N_0} \sum_{\mathbf{q},\nu} \delta(\omega-\omega_{\mathbf{q},\nu}) \frac{\gamma_{\mathbf{q},\nu}}{\omega_{\mathbf{q},\nu}} ~. \label{a2F}
\end{align}
The characteristic phonon energy scale $\omega_{\rm log}$ is defined as
\begin{align}
\omega_{\mathrm{log}} = \exp\Big( \frac{2}{\lambda} \int_0^{\infty} \frac{\mathrm{d}\omega}{\omega} \alpha^2F(\omega) \log(\omega) \Big) ~.
\end{align}

\begin{figure}[h!]
	\includegraphics[width=0.7709\linewidth]{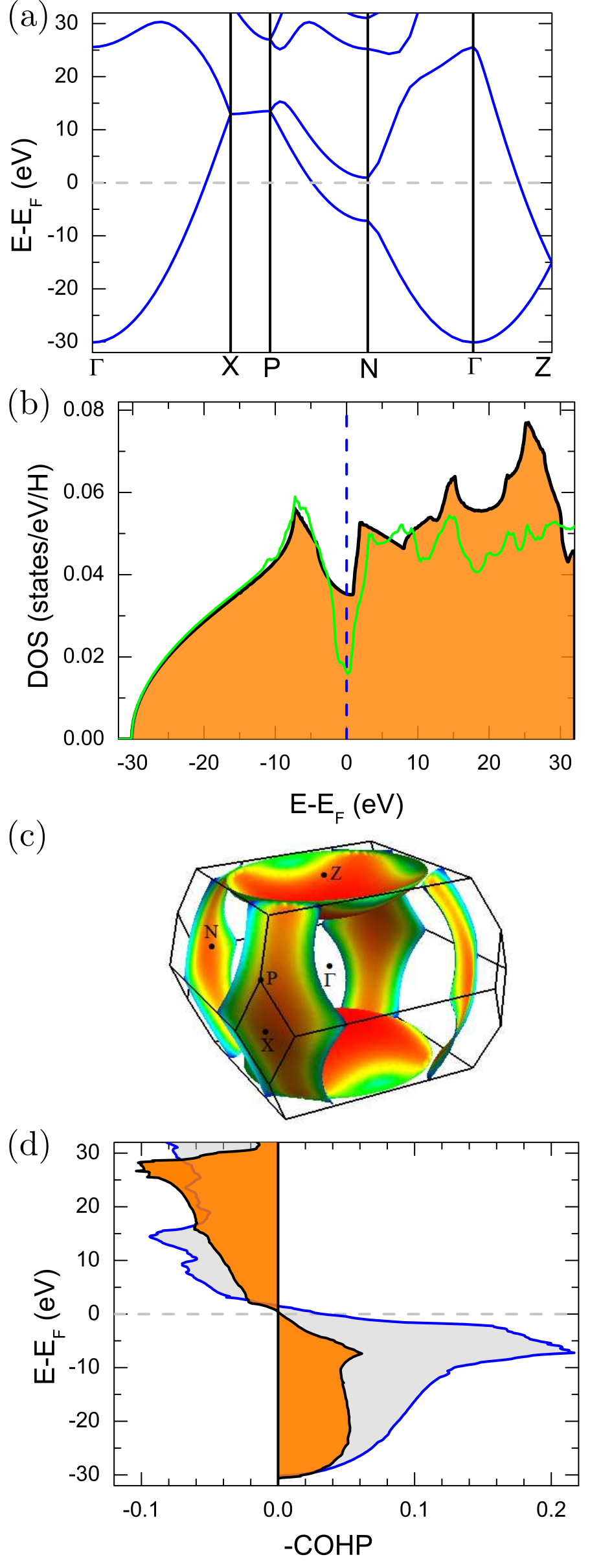}
	\caption{\textit{Ab initio} calculated electronic properties of metallic hydrogen at 400 GPa. (a) Electronic band structure along the high symmetry directions of the BZ. (b) Electronic density of states (DOS) function, shown in orange color. The light green curve show the DOS results of the $Cmca$-8 phase of molecular hydrogen {\cite{Pickard2007}}.  (c) Computed Fermi-surface sheets, colored according to the Fermi velocity values,  starting with blue (lowest velocities) -- light green (medium range velocities) -- red colors (highest velocities),  (d) The COHP functions. The orange shaded area shows the result for atomic hydrogen and the blue curve (grey shaded) shows the result for the $Cmca$-8 phase of molecular hydrogen. Negative COHP values represent bonding interactions and positive COHP values represent anti-bonding interactions.}
\end {figure}

\section{Results}\label{scResults}

We begin by calculating the electronic properties of metallic hydrogen in the optimized $I4_1/amd$ structure. The results of our electronic structure calculations at a pressure of 400 GPa are presented in Fig.\ 1. Figure 1(a) shows the computed electronic bandstructure plotted along high-symmetry directions in the {BZ}. The electronic states are highly dispersive forming very wide bands which reflects their nearly free electron nature. Two bands cross the Fermi level and are responsible for metallicity. One band corresponds to the bonding $s$-orbital and {the} other {one} to {the} antibonding $s$-orbital. The antibonding state is mostly unoccupied and crosses along the $\Gamma -$Z direction. Our calculated electron bandstructure agrees well with previously reported results \cite{Borinaga2016,Degtyarenko2016,Kudryashov2017}.

The electronic density of states (DOS), shown in Fig.\ 1(b) with the orange color, is also consistent with the free electron behavior, being nearly parabolic below to Fermi level. Note that the DOS value at the Fermi level is higher than that of metallic molecular hydrogen in the $Cmca$-8 phase \cite{Pickard2007}, shown in green. Figure 1(c) shows the calculated 3D Fermi surface of atomic hydrogen metal at 400 GPa, which consists of two sheets corresponding to the bonding and antibonding states. The bonding state leads to ribbon-like hole Fermi surface sheets and the antibonding state leads to a concave lens shaped electron Fermi sheet at Z
(the Fermi surfaces were rendered using FermiSurfer software \cite{Kawamura2019}). The sheets in Fig.\ 1(c) are colored according to the values of the Fermi velocities; the high Fermi velocities correspond to the free electron nature. The covalent character of the H-H bonds is investigated by calculating the crystal orbital Hamiltonian populations (COHP) functions \cite{Dronskowski1993,Grechnev2003,Steinberg2018,Andersen,Verma2018} which counts the population of wavefunctions on two atomic orbitals of a pair of atoms (shown in Fig.\ 1(d)). In a given energy window, negative values of COHP describe bonding interactions whereas positive values of COHP describe anti-bonding interactions. This analysis shows that the overlap of nearest hydrogen states below the Fermi level are bonding states. The H-H bond in molecular phase (grey shaded area) has stronger covalent character than that of atomic phase. The integrated COHP values (computed with the code of Ref.\ \cite{Andersen}) are 1.30 and 3.27 eV/H-H for atomic and molecular phases, respectively.

\FloatBarrier

The computed phonon dispersions of metallic hydrogen at 400 GPa (not shown) are very similar to the previously reported phonon dispersions \cite{Borinaga2016}.

After this we turn our attention to the superconducting properties of metallic hydrogen. In Fig.\,\ref{lambdaA2F}(a) we show our convergence study of {the global electron-phonon coupling strength} $\lambda$, as obtained from Eq.\,(\ref{lambda_sum}), as function of smearing $\sigma$, which is used by Quantum Espresso to compute the electron-phonon coupling coefficients $\lambda_{\mathbf{q},\nu}$. The results show good convergence for $\sigma\gtrsim0.04\,\mathrm{Ry}$, and therefore this value of $\sigma$ will be used from here on. The converged value of $\lambda$ is  {2.32, which} is in agreement with the {$\lambda=2.08$} computed for molecular hydrogen at 450 GPa \cite{Cudazzo2008}. The coupling coefficient is also consistent with values for H$_3$S at 200 GPa (2.19) \cite{Duan2014}, LaH$_{10}$ at 250 GPa (2.29) \cite{Liu2017}, and YH$_{10}$ at 400 GPa (2.41) \cite{Peng2017}.
\begin{figure}[t!]
	\centering
	\includegraphics[width=1\linewidth]{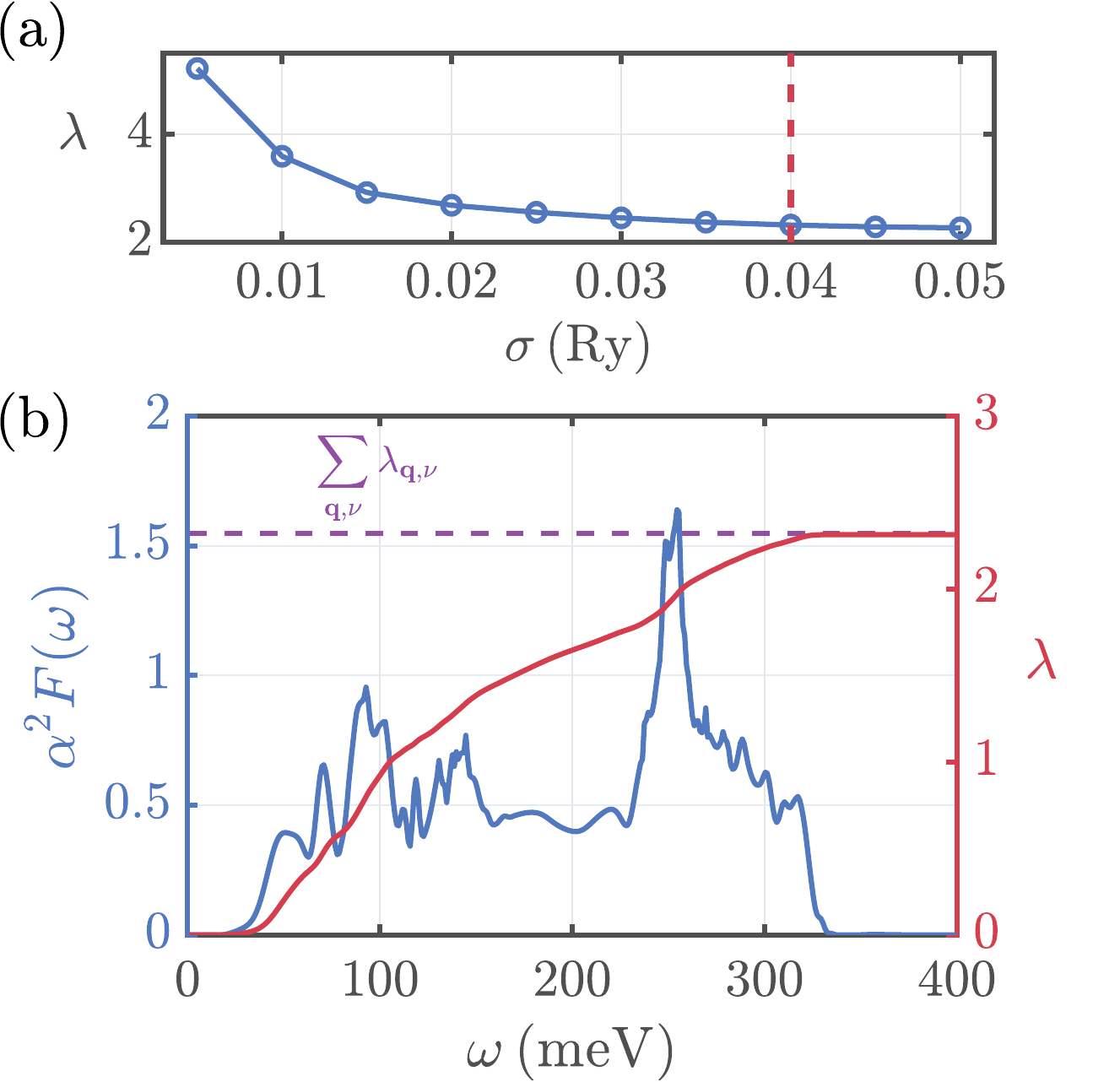}
	\caption{(a) Global coupling constant as function of broadening. (b) Frequency dependent Eliashberg function for a broadening value of 0.04 Ry (blue curve). The red curve shows the cumulative coupling strength calculated from Eq.\,(\ref{lambda_int}). The purple dashed line represents $\lambda$ as obtained from Eq.\,(\ref{lambda_sum}).}\label{lambdaA2F}
\end{figure}

Now we turn to the analysis of how individual phonon modes contribute to the electron-phonon couplings. For this we {calculate} the Eliashberg function $\alpha^2F(\omega)$,  shown in Fig.\,\ref{lambdaA2F}(b) in blue. The most prominent contributions appear at $\omega\sim100\,\mathrm{meV}$ and $\omega\sim250\,\mathrm{meV}$.
This is further emphasized by the red curve representing the cumulative electron-phonon coupling as calculated from Eq.\,(\ref{lambda_int}). The aforementioned frequencies lead to the steepest increase in $\lambda$ with $\omega$. As crosscheck, we calculated the total electron-phonon coupling using Eq.\,(\ref{lambda_sum}), shown in dashed purple. Both calculations yield identical values of $\lambda$. 
\begin{figure}[h!]
	\centering
	\includegraphics[width=1\linewidth]{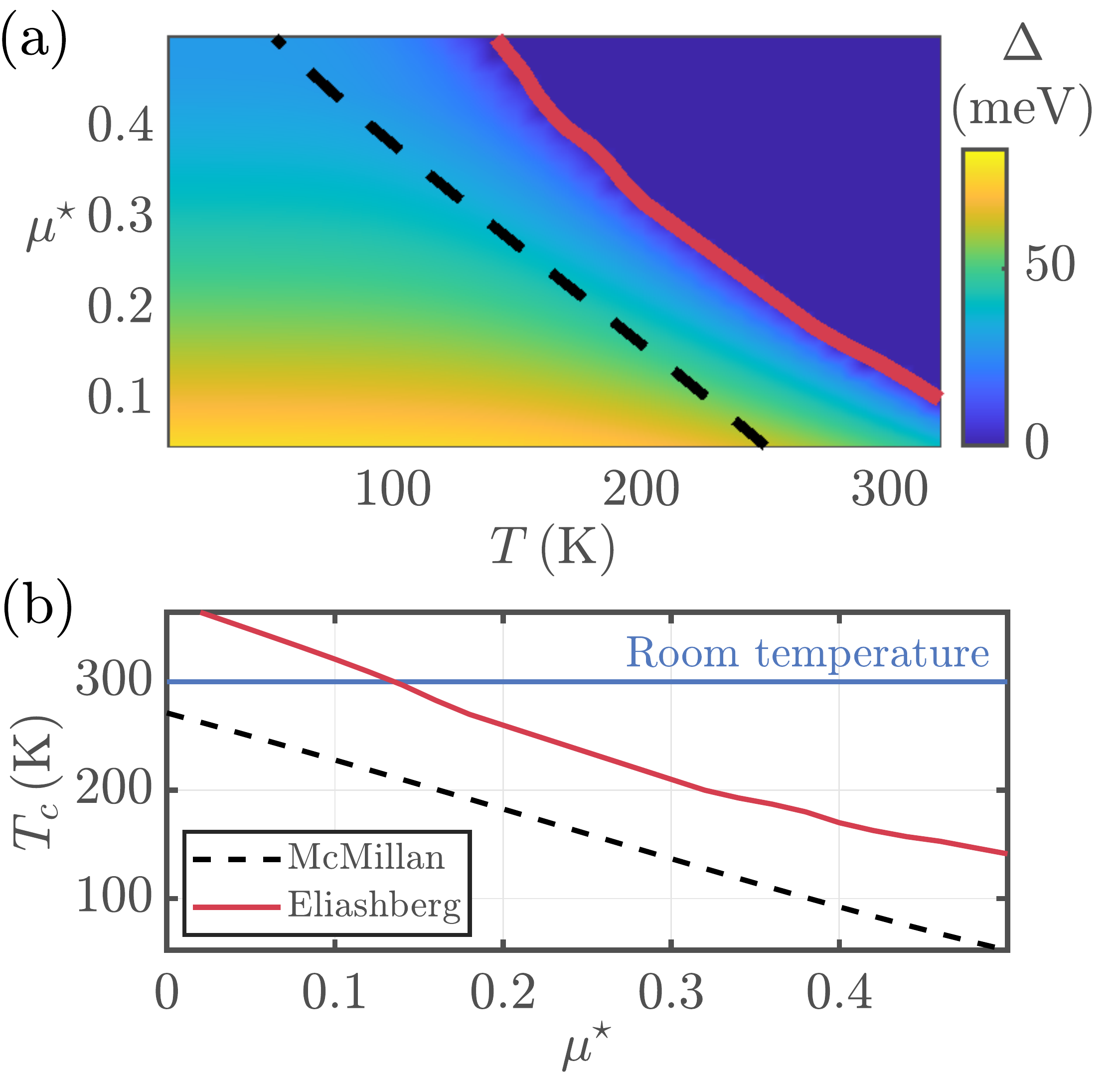}
	\caption{(a) Self-consistently computed maximum superconducting gap for atomic hydrogen at 400 GPa as function of $T$ and screened Coulomb potential $\mu^{\star}$. The critical temperatures according to Eliashberg theory and the modified McMillan equation are drawn in solid red and dashed black lines, respectively. (b) Calculated critical temperature $T_c$ plotted against $\mu^{\star}$.}\label{gaptc}
\end{figure}

Next we solve the Eliashberg equations as function of $T$ and pseudopotential $\mu^{\star}$, using the first-principles input computed for $\sigma=0.04\,\mathrm{Ry}$. We show the result for the maximum superconducting gap $\Delta=\mathrm{max}_{\mathbf{k}}\,\Delta_{\mathbf{k},m=0}$ in Fig.\,\ref{gaptc}(a).
In solid red we indicate the onset of superconductivity, hence the critical temperatures. Above we mention another recipe of calculating $T_c$ by means of the modified McMillan equation; the outcome is plotted as dashed black line. We make the $\mu^{\star}$-dependence of $T_c$ explicit in Fig.\,\ref{gaptc}(b), where we show the $T_c$ corresponding to room temperature in solid blue. As apparent, when using the modified McMillan equation we underestimate the critical temperature, in comparison to the solution of the more accurate Eliashberg equations, for all values of $\mu^{\star}$. The dashed black line stays below room temperature even in the complete absence of pair-breaking Coulomb repulsion. The red solid line, on the other hand, predicts room temperature superconductivity for values of $\mu^{\star}$ up to $\sim0.14$. 

We now turn to the question about the most significant phonon branches. The irreducible representations in this system are $B_{1g}$ (one mode), $E_g$ (two modes), $A_{2u}$ (one mode), and $E_u$ (two modes). We split $\lambda_{\mathbf{q},\nu}$, $\gamma_{\mathbf{q},\nu}$, and $\omega_{\mathbf{q},\nu}$ according to these subsets, and repeat the calculation of $\lambda$ and $\alpha^2F(\omega)$, respectively via Eq.\,(\ref{lambda_sum}) and Eq.\,(\ref{a2F}). The relative contribution to the electron-phonon coupling due to the different phonon modes is shown as inset in Fig.\,\ref{splita2F}. In the main graph, we plot the partial Eliashberg functions arising from each irreducible representation. Concerning $\alpha^2F(\omega)$, we clearly see that each subset of phonon modes contributes mainly in a respective characteristic frequency range. As for the magnitude of $\lambda$, the largest (smallest) contributions are due to $B_{1g}$ ($A_{2u}$), while $E_u$ and $E_g$ are on a comparable intermediate level.

\begin{figure}[t!]
	\centering
	\includegraphics[width=1\linewidth]{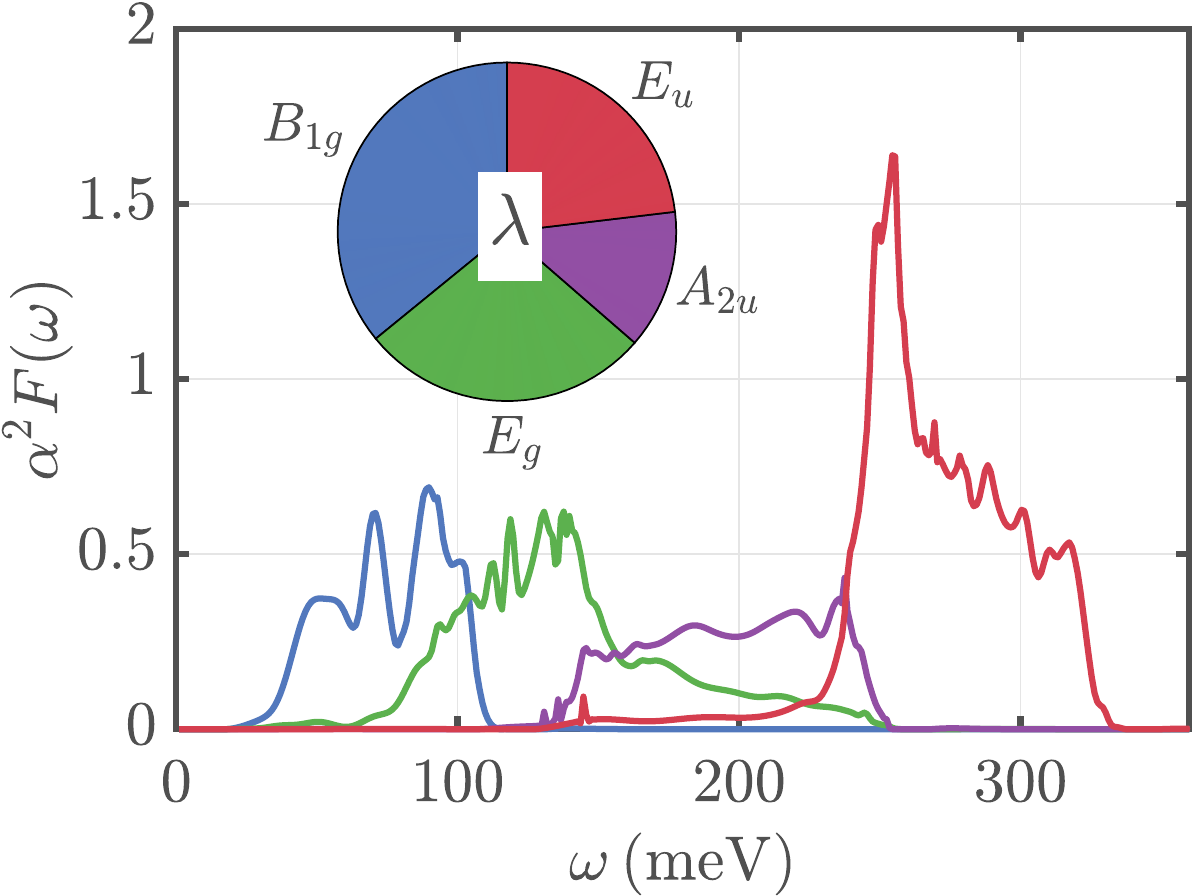}
	\caption{Frequency dependent Eliashberg function of atomic hydrogen at 400 GPa produced by the four different irreducible representations. The inset shows the contributions of $B_{1g}$, $E_g$, $A_{2u}$ and $E_u$ to the global electron-phonon coupling constant. The same colors are used for the main graph and the inset.}\label{splita2F}
\end{figure}
\begin{figure}[b!]
	\centering
	\includegraphics[width=1\linewidth]{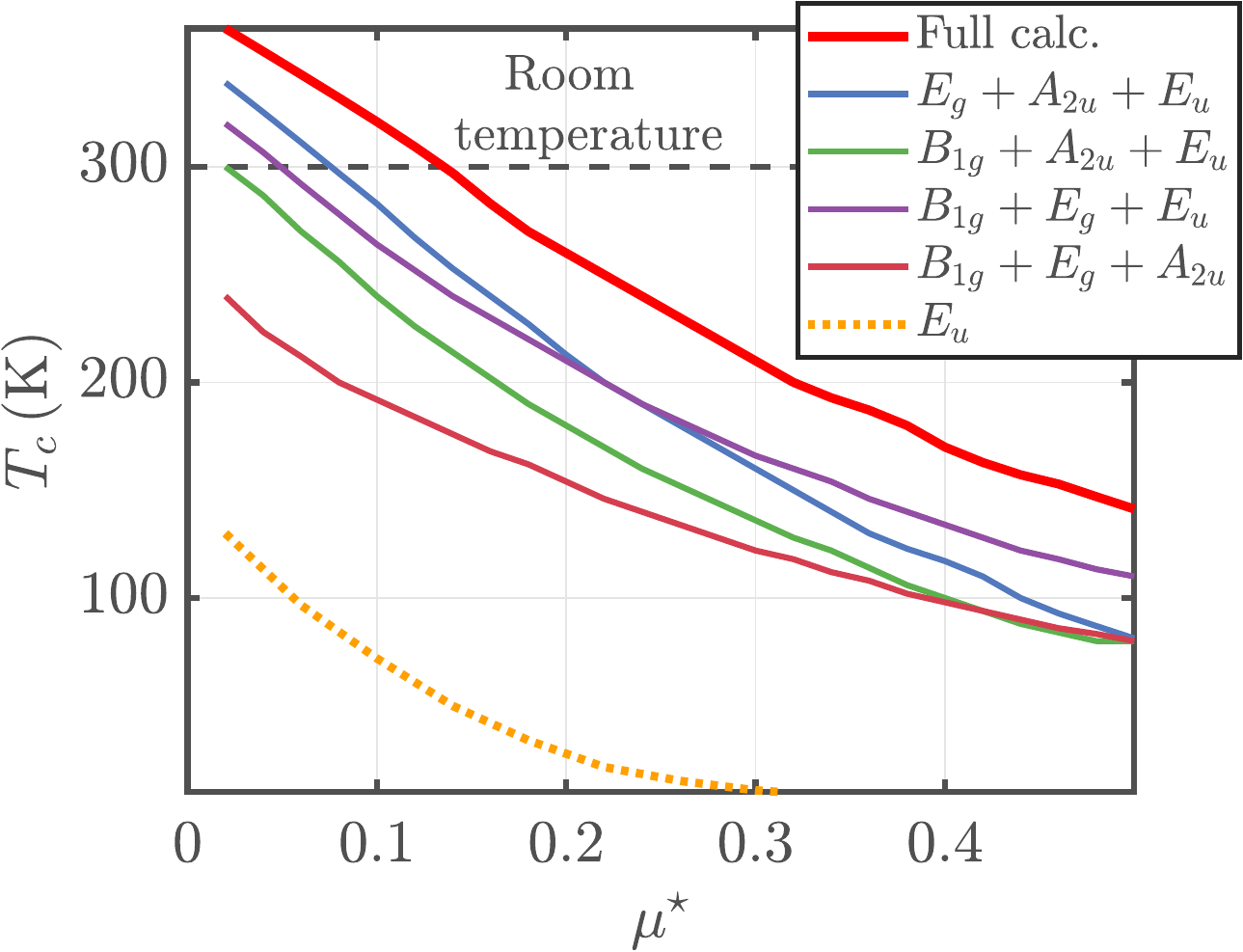}
	\caption{Calculated transition temperatures as function of the Coulomb pseudopotential. The red thick line represents our result for including all modes in the system. The remaining four solid lines are found by neglecting one subset of phonon modes at a time, see legend. The dotted yellow curve is calculated from the $E_u$ modes only.}\label{splitTc}
\end{figure}

We want to examine how the different phonon modes affect the superconducting transition temperature. The $\mu^{\star}$ dependent result for $T_c$ as obtained from the full calculation is shown in Fig.\,\ref{splitTc} as fat, light red curve. The calculations are now repeated by selectively leaving out one particular subset of phonon modes. For example, the blue line in Fig.\,\ref{splitTc} is found by taking into account only the $E_g$, $A_{2u}$ and $E_u$ irreducible representations, i.e., neglecting any influence due to $B_{1g}$.  
From this we observe that the smallest decrease in $T_c$ is found when leaving out either the $B_{1g}$ or the $A_{2u}$ modes, hence their significance for superconductivity is comparatively minor. The largest loss in $T_c$ is found when excluding the $E_u$ modes, see the dark red curve. To investigate this representation closer, we perform calculations with the $E_u$ modes only, shown as dotted yellow curve in Fig.\,\ref{splitTc}, and find a maximum $T_c\sim140\,\mathrm{K}$. This contribution is significantly larger than found for any other isolated irreducible representation (not shown). Hence we conclude that phonon modes belonging to the $E_u$ representation are most important for the high-temperature superconducting state.

\begin{figure}[b!]
	\centering
	\includegraphics[width=1\linewidth]{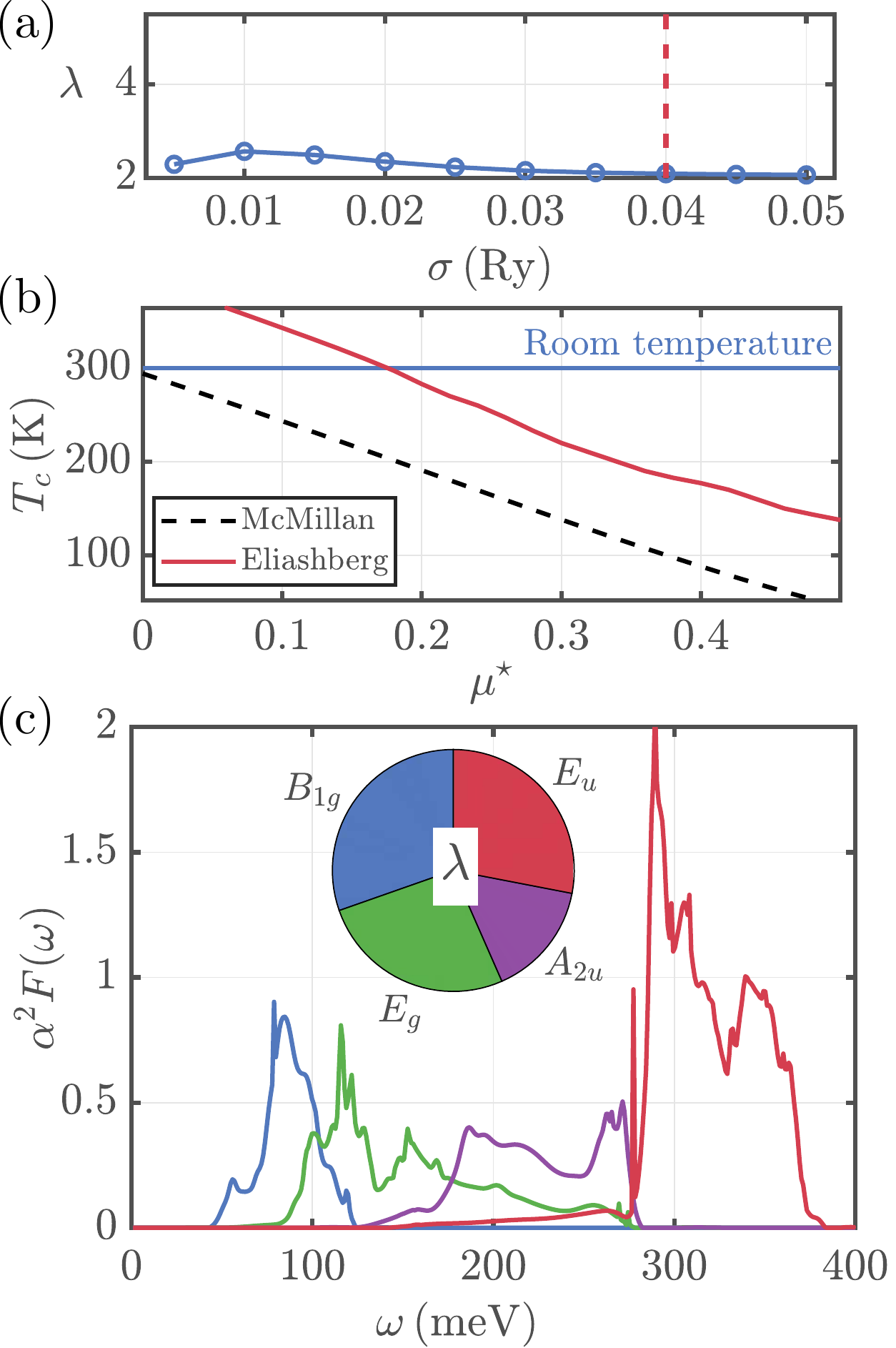}
	\caption{(a) Computed global coupling constant $\lambda$ as function of broadening for atomic hydrogen at a pressure of 600 GPa. (b) The critical temperature versus Coulomb pseudopotential $\mu^{\star}$ computed with Eliashberg theory and with the modified McMillan equation. (c) As Fig.\,\ref{splita2F}, but for atomic hydrogen at {a pressure of 600} GPa.}\label{lambdaTc600GPa}
\end{figure}

We performed similar calculations for atomic hydrogen at 600 GPa. As shown in Fig.\ \ref{lambdaTc600GPa}(a), we find a slight decrease in the value of electron-phonon coupling, $\lambda\,=\,2.09$, but overall,  as can be seen in Fig.\ \ref{lambdaTc600GPa}(b)}, $T_c$ increases slightly. Although this behavior may seem counter-intuitive, it can be explained by the way different phonon modes contribute to superconductivity. Despite the small decrease in the total electron-phonon coupling, we now find an increased coupling to $E_u$ symmetry modes, as illustrated in the inset of Fig.\ \ref{lambdaTc600GPa}(c), which leads to the increase in transition temperature. Hence, this underlines the dominant contribution stemming from the $E_u$ phonon modes. \\

\section{CONCLUSIONS}
In summary, we have reported a detailed analysis of the superconducting properties of metallic atomic hydrogen under high pressure conditions. To this end, we solved the anisotropic Migdal-Eliashberg equations, in combination  with first-principles input results for the electron energies, phonons, and electron-phonon couplings. Our calculations show that, although the H-H covalent bond has weakened in atomic hydrogen metal as compared to the molecular hydrogen phase, it still has a substantial amount of covalent character. Further we find that metallic atomic hydrogen exhibits above room temperature superconductivity for reasonable values of the screened Coulomb pseudopotential  $\mu^{\star}$. Analyzing which modes contribute most, we find that the high transition temperature is mainly due to the $E_u$ phonon modes. The critical transition temperature shows only a slight increase with pressure.

\begin{acknowledgments}
A.K.V.\ and P.M.\ acknowledge the support of ANUPAM supercomputing facility of BARC. 
F.S., A.A., and P.M.O.\ acknowledge support by the Swedish Research Council (VR), the R{\"o}ntgen-{\AA}ngstr{\"o}m Cluster, the Knut and Alice Wallenberg Foundation (No.\ 2015.0060), and the Swedish National Infrastructure for Computing (SNIC).
\end{acknowledgments}

\bibliographystyle{apsrev4-1}

\end{document}